\documentclass[aps,prl,showpacs,twocolumn,superscriptaddress]{revtex4}
\usepackage{epsfig}
\usepackage{amsfonts}
\usepackage{amsmath}
\usepackage[T1]{fontenc}

\newcommand{\beq}{\begin{equation}}
\newcommand{\eeq}{\end{equation}}
\newcommand{\bea}{\begin{eqnarray}}
\newcommand{\eea}{\end{eqnarray}}
\newcommand{\nn}{\nonumber}

\newcommand{\al}{\alpha}

\begin{document}
\title{Disordered one-dimensional Bose-Fermi mixtures: The Bose-Fermi glass}
\author{Fran\c cois Cr\'epin}
\affiliation{Laboratoire de Physique des Solides, CNRS UMR-8502 Universit\'e Paris Sud, 91405 Orsay Cedex, France}
\author{Gergely Zar\'and}
\affiliation{Budapest University of Technology and Economics, H-1521 Budapest, Hungary}
\author{Pascal Simon}
\affiliation{Laboratoire de Physique des Solides, CNRS UMR-8502 Universit\'e Paris Sud, 91405 Orsay Cedex, France}

\date{\today}

\begin{abstract}
We analyze an interacting Bose-Fermi mixture in a 1D disordered potential using a combination of renormalization group and variational methods. We obtain the complete phase diagram {in the incommensurate case}
as a function of bosonic and inter-species interaction strengths, in the weak disorder limit.
We find that the system is characterized by several phase transitions
between superfluid and various glassy insulating states, including a
new Bose-Fermi glass phase, where both species are coupled and
localized. We show that the dynamical structure factor, as measured
through Bragg scattering experiments, can distinguish between the
various localized phases and probe their dynamics. 
\end{abstract}

\pacs{67.60.Fp, 67.85.Pq, 71.10.Pm, 71.23.-k, 71.55.Jv}

\maketitle

\emph{Introduction.}
Spectacular developments of experimental techniques in the field of
ultra-cold atoms have opened a new way in the study of strongly
correlated systems \cite{bloch1}. Experimentalists are now able to create
optical lattices with highly tunable
parameters, and realize various  models and quantum phase
transitions.
In Ref.~\cite{bloch2}, e.g.,  superfluid bosons were driven to a Mott
 insulating state, by tuning the on-site
interactions of the Hubbard model, and the  incompressibility
of this state has also been
demonstrated~\cite{esslinger}. The Mott
transition was also realized in Fermi systems \cite{bloch3},
where other interesting phenomena  such
as pairing with spin imbalance \cite{hulet2,ketterle2} 
or the BCS-BEC crossover \cite{bloch1} were also
investigated.   

Having more and more control over conventional systems,
experimentalists and theorists now turn to the study
of more complicated ones.
On the one hand, a lot of attention is devoted to
{\em multicomponent} systems such as Bose-Fermi
mixtures~\cite{salomon,ketterle3,bongs,esslinger2,bloch4}, or
three component systems~\cite{3comp}, which offer the possibility
of realizing new phases of matter, such as supersolids,
color superconductivity~\cite{colorSC}, or 'baryonic'
phases~\cite{zarand}.  On the other hand,
the creation of {\em disorder} using speckle lasers or
incommensurate laser beams in trapped systems
paved the road to create quantum glasses and Anderson
insulators in cold atomic systems~\cite{aspect1,inguscio1}.

\begin{figure}
\includegraphics[width=7.5cm]{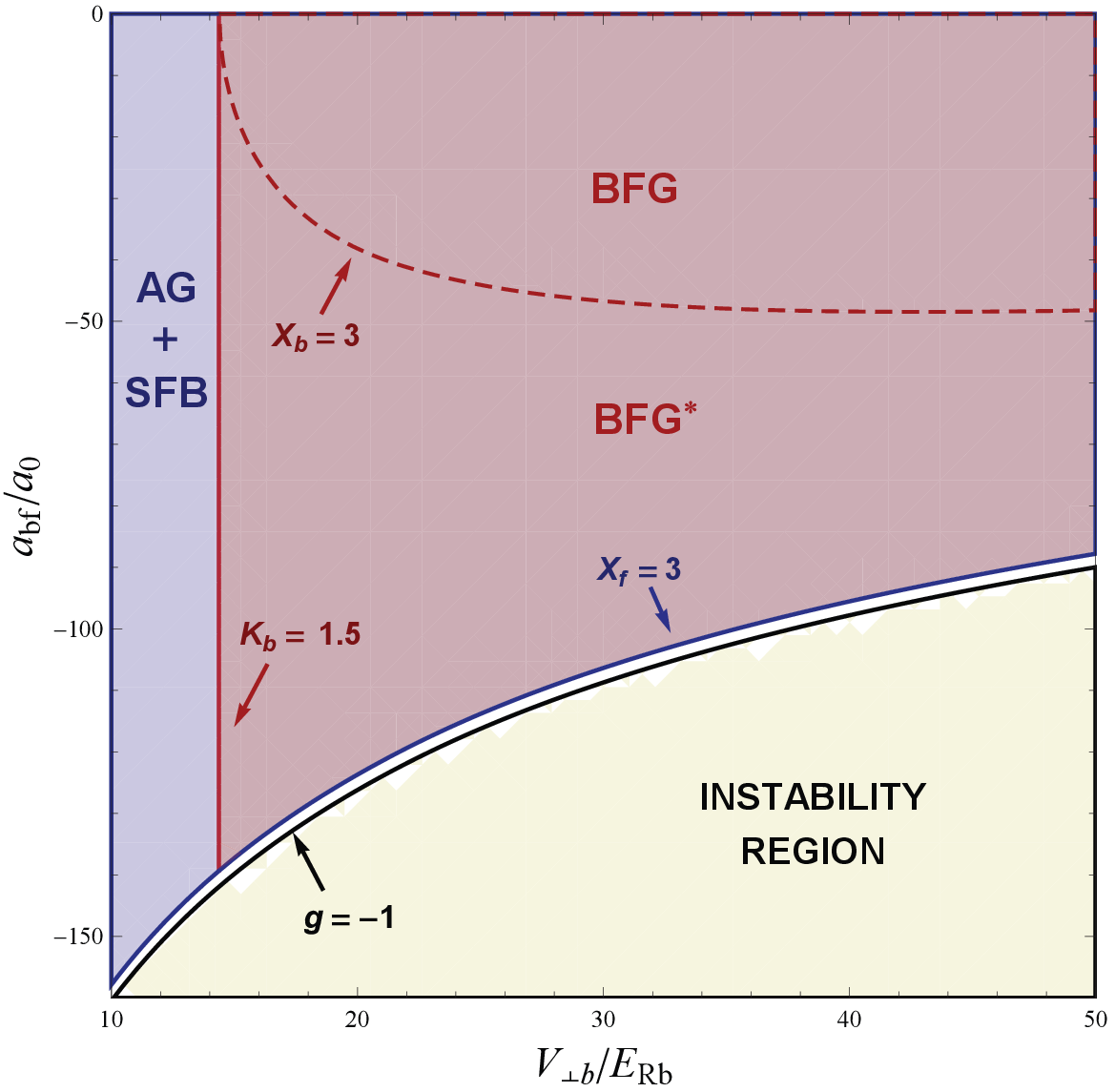}
\caption{
Phase diagram of a 1D $^{87}$Rb-$^{40}$K Bose-Fermi mixture, in the
weak disorder limit. We consider an array of tubes,
created with  lasers of wavelength  $\lambda$ = 1064 nm, which corresponds
to a (2D) lattice constant $d=\lambda/2$.
We take  $a_{bb}=100 a_0$ as 
Bose-Bose scattering length, while  the Bose-Fermi scattering length, $a_{bf}$, is
tuned using a Feshbach resonance. The one.dimensional densities are
chosen to be $\rho_f d = 0.3 $ and $\rho_b d= 0.1 $.   
The recoil energy is $E_{R,b}=h^2 \lambda^{-2}/(2m_{Rb})$, while
$V_{\perp b}$ is the transverse confining potential creating the 1D
tubes. { The Bose-Bose interaction $U_b$ increases with $V_{\perp b}$} \cite{epaps}. Four phases and the region { of instability of the Luttinger liquid theory}  are shown.
BFG: Bose-Fermi glass,  BFG$^*$ (BFG
with an extremely large bosonic localization length), AG+SFB: Anderson
Glass + Superfluid Bosons, LL: Luttinger  liquid ({above the instability region}) . 
}
\label{Fig:phase_diag_exp}
\end{figure}

It is well known that impurities can lead to
the localization of electronic wave 
functions. While interactions are known to be  important in the localized
phase, taking them into account is quite a difficult task.
In one dimension, this question has been addressed
both for fermions \cite{Giamarchi_anderson} and for bosons
\cite{Giamarchi_anderson,Fisher}. For spinless
fermions, sufficiently attractive interactions trigger a 
transition from the Anderson insulating phase to a p-wave
superconducting phase. For bosons, a disorder-induced  insulating
phase appears for sufficiently strong interactions: nicknamed Bose
glass,  it has no gap, it is insulating, and is
compressible~\cite{Fisher}.  The Bose glass disappears at
 sufficiently small repulsive
interactions, where the system enters a superfluid phase. On the other hand,
for even smaller  interactions bosons should undergo Anderson localization and
condense in the lowest eigenstate
\cite{Giamarchi_anderson,Fisher}. The Bose
glass~\cite{Inguscio_disorder} and the Anderson glass~\cite{aspect1}
phases were now observed experimentally.


In the present paper we wish study,
how  the most basic two-component system, the Bose-Fermi mixture (BFM)
is affected by the presence of disorder.
This is not only important in view of the current experimental interest
\cite{salomon,ketterle3,bongs,esslinger2,bloch4},
but it is a challenging fundamental question on its own, too,
especially in the limit of strong interactions.
In particular, one may wonder whether disorder ``decouples'' both species or
whether the localization of one species triggers the
localization of the other species because of interactions. This issue
has been discussed in a general context in
Ref. \cite{Lewenstein}, where a variety of exotic situations were
considered.  In this letter we focus on a disordered one-dimensional
(1D) BFM,  where interactions are particularly relevant.
One-dimensional  BFMs in the absence of disorder
have been investigated by a variety of
analytical~\cite{cazalilla1,polaron1}
and numerical \cite{Hebert1} techniques. 
Here, we follow Refs.~\cite{Giamarchi_anderson,Fisher} and use  bosonization
to extract the low-energy properties of the BFM  in a disordered potential. 
We focus on the case where the bosonic and fermionic densities are not 
commensurate either with the lattice constant or with each other. 
Our main results are summarized for a $^{87}$Rb-$^{40}$K
mixture  in Fig.~\ref{Fig:phase_diag_exp}.
We find two different glassy phases as well as a Luttinger
liquid phase and a region {where the Luttinger liquid theory breaks down (see caption).}

\emph{Model.} We consider a 1D mixture of bosons and spinless fermions
described by the following Hamiltonian,
$H = H_b + H_f + H_{bf} + H_{\rm dis}$,
where
\bea
H_{\alpha} &=&   \int dx\;
\Bigl[
- \psi_{\alpha}^\dag(x)\frac{\partial_x^2}{2 m_{\alpha}} \psi_{\alpha}(x)
+ \frac{U_{\alpha}}{2} :\rho^2_{\alpha}(x):
\Bigr],   
\nn
\\
H_{bf} &=&  U_{bf}  \int dx \;    \delta\rho_{b}(x) \delta\rho_{f}(x),  \\
H_{\rm dis} &=&  \int dx \;  \Bigl(
V_{b}(x)  \rho_{b}(x) + V_{f}(x)  \rho_{f}(x)
\Bigr)
\;.
\eea
Here $\psi_{\alpha}$ ($\alpha=b,f$)  denote the Bose/Fermi fields,
$\rho_{\alpha}(x)= \psi_{\alpha}^\dagger(x)\psi_{\alpha}(x)$ their density,
and $m_\alpha$ the mass of the bosons/fermions. The
symbol $:\dots:$ stands for normal ordering. The interaction between
bosons must be repulsive for stability reasons, $U_b>0$, while for
spinless  fermions with a local interaction, $U_f\equiv 0$.
Bosons and fermions can  have either repulsive ($U_{bf}>0$) or attractive ($U_{bf}<0$) interactions. The  random potentials, $V_{f/b}(x)$ describe the effects
of disorder. For simplicity, here we assume them to be Gaussian
with zero mean values, and  spatially uncorrelated, {\it i.e.},
 $\overline{V_{\al}(x)V_{\al}(x') } = D_{\al} \delta (x-x')$
\footnote{Overlining a quantity denotes disorder average.}. 
Experimentally, there is only one disorder potential, for
instance a laser speckle. As a result, $V_f$ and $V_b$ are
correlated~\footnote{Integrating out high momentum fermions (bosons)
also generates $V_b$ ($V_f$) through $U_{b/f}$, even if the
disorder couples only to fermions (bosons).}.
Nevertheless, $\rho_b$ being incommensurate with
$\rho_f$, we find that for long wavelength excitations $V_b$ and
$V_f$ act as if they were independent~\cite{future_paper}. In the following,
we therefore assume   $\overline{V_{f}(x)V_{b}(x')} = 0$.

In 1D, long wavelength excitations  are essentially density fluctuations,
and the densities $\varrho_{f/b}$ and the  fields $\psi_{f/b}$ can be
"bosonized", i.e., 
represented in terms of bosonic  phases, $\phi_{f/b}$~\cite{Haldane}.
Then $H_f$ and $H_b$ take on simple quadratic forms
\beq
H_{\alpha} = \frac{v_\alpha}{2\pi}\int dx \left[K_\alpha\left( \partial_x \theta_\alpha(x)\right)^2 + \frac{1}{K_\alpha} \left( \partial_x \phi_\alpha(x)\right)^2 \right],
\eeq
with $\phi_\al(x)$, $\theta_\al(x)$  denoting  density and phase quantum fields, obeying
 $\left[ \phi_{\al}(x'), \partial_x \theta_{\al}(x) \right] = i\pi\delta (x-x')$.
The prefactor $v_b$ can be viewed as the sound velocity for Bogoliubov phonons
in a quasi-condensate, while $v_f$ is the Fermi velocity. 
The dimensionless Luttinger parameters,  $K_f$ and $K_b$, characterize the strength of
interactions \cite{cazalilla2}:
In our case of  non-interacting fermions, $K_f = 1$.
For bosons, $K_b = 1$ in the hardcore limit, $K_b > 1$ for softer
repulsive interactions, and $K_b \rightarrow \infty$ for
non-interacting bosons. 
Keeping in mind that $\varrho_b$ is incommensurate with  $\varrho_f$,
backward scattering of bosons on fermions is irrelevant, and at low energies $H_{bf}$ 
can be approximated as
\beq
H_{bf} = \frac{U_{bf}}{\pi^2}  \int dx \ \partial_x \phi_b(x) \partial_x \phi_f(x).
\eeq
In the same way, disorder induces backward scattering and forward scattering.
However, the forward scattering can be gauged away
(without affecting current-current and superfluid correlations),
and only backward scattering responsible for localization
remains~\cite{Giamarchi_anderson,future_paper},
\beq
H^{\alpha}_{dis}=\int dx \left[
\rho_\alpha \xi_\alpha(x)e^{- i2\phi_\alpha(x)} + h.c. \right],
\eeq
with  $\overline{\xi_\alpha(x)\xi^*_{\alpha}(x')} = D_\alpha \;\delta(x-x')$.

\emph{Renormalization group} (RG). To perform the RG
analysis, we use the replica trick. We introduce $n$
replicas of the system, average over disorder, and then take
$n \to0$.  The replicated action then reads $S_{\rm rep} = S_0 +
S^{\rm rep}_{\rm dis}$,  with
\begin{widetext}
\begin{eqnarray}
S_0 &=& \sum_{a=1}^n \sum_{\alpha=f,b}\frac{1}{2\pi K_\alpha}\int dx d\tau
\left[\frac{1}{v_\alpha}\left( \partial_\tau \phi^a_\alpha\right)^2 + v_\alpha \left(
  \partial_x \phi^a_\alpha\right)^2 \right] 
+ \sum_{a=1}^n \frac{U_{bf}}{\pi^2}\int dx d\tau \ \partial_x\phi^a_f \partial_x \phi^a_b,
\label{eq:s0}\\
\label{Srep} S^{\rm rep}_{\rm dis} &=&  - D_f \rho_f^2 \sum_{a,b}\int dx d\tau d\tau ' \cos(2\phi_f^a(x,\tau) - 2\phi_f^b(x,\tau ')) 
- D_b \rho_b^2 \sum_{a,b}\int dx d\tau d\tau ' \cos(2\phi_b^a(x,\tau) - 2\phi_b^b(x,\tau ')).
\end{eqnarray}
\end{widetext}
Integrating out high momentum degrees of freedom
we find the following flow equations to lowest order,
\begin{eqnarray}
\frac{d\tilde{D}_\al}{dl} &=& (3-X_\al)\tilde{D}_\al(l)\;, \label{eq:D} \\
\frac{d}{dl}(K_\al v_\al) &=& - 2 K_\al(l) v_\al(l)
\tilde{D}_\al(l) \ \mathcal{C}_\al
\label{K_scaling}
\end{eqnarray}
while  $\frac{d}{dl} (K_\al/v_\al) = 0$.
Here we have defined dimensionless variables for the disorder,
$\displaystyle \tilde{D}_\alpha =  \frac{2}{\pi
  \Lambda_0}\frac{K_\alpha^2}{v_\alpha^2} D_\alpha$, and parametrized
the high-energy cut-off as $\Lambda = \Lambda_0 e^{-l}$.
The  $\mathcal{C}_\al$ are numbers coming from our RG scheme. The
anomalous dimensions, $X_{f/b}$, of the disorder operators  can
be obtained from the diagonalization of $S_0$ and read:
\begin{equation}\label{eq:xf}
X_f =  \frac{2K_f(1+t\sqrt{1-g^2})}{\sqrt{1-g^2}\sqrt{1+2t\sqrt{1-g^2}+t^2}},
\end{equation}
 where $t=v_f/v_b$ and $g=U_{bf}/\pi \sqrt{K_f K_b/(v_f v_b)}$ is a
dimensionless parameter. The dimension
$X_b$ is obtained by changing $t\to 1/t$ and
$K_f$ into $K_b$ in Eq (\ref{eq:xf}).

For uncoupled species ($g=0$), $X_f=2K_f$ and $X_b=2K_b$. Thus
spinless fermions (bosons) are localized when $K_f<3/2$ ($K_b<3/2$)
\cite{Giamarchi_anderson}. At weak disorder we can neglect
the feed-back, Eq.~\eqref{K_scaling}, and follow, e.g.
Ref.~\onlinecite{cazalilla2}, to relate $K_f$, $K_b$, $t$, $g$ to experimental
parameters, and construct the phase diagram of Fig.~\ref{Fig:phase_diag_exp}.
The thick solid blue line denotes $X_f = 3$, while the dashed red line
indicates $X_b = 3$. Naively, one would expect these lines to separate
three phases:  In the Bose-Fermi glass phase (BFG) $X_b<3$ and
$X_f<3$, both species are pinned by disorder, and
fermionic and bosonic excitations are localized
over the localization lengths, $\xi_f$ and $\xi_b$, respectively.
In the  Luttinger liquid  (LL) phase $X_b>3$ and $X_f>3$, disorder is
thus irrelevant, and   both species are superfluid. In the regime,
$X_b>3$ and $X_f<3$ one would naively predict a phase with localized
fermions forming an Anderson glass (AG)
and superfluid bosons (SFB). In this regime, however, some care
needs be taken: since fermions are localized,  fermionic density fluctuations
become "gapped" at a length scale
$\Lambda\sim 1/\xi_f$. Below this scale bosons
interact with their bare interaction. Thus the RG equations become
\beq
\frac{d \textrm{log} \tilde{D}_b}{dl} =
\left\{
\begin{array}{ll}
3 - X_b     & \mbox{ if $\Lambda\gg1/\xi_f$}\;\\
3-2K_b     & \mbox{ if $\Lambda\ll1/\xi_f$}.
\end{array}
\label{Db_scaling}
\right.
\eeq
As a consequence, the region $X_b>3$ and $X_f<3$ is divided into
two phases. In the region, $X_f<3$ and $K_b>3/2$, we obtain
an AG+SFB phase. However, in the region $X_f<3$ and $K_b<3/2$ bosons
are ultimately localized on a very large length scale, and we find
a Bose-Fermi glass phase (BFG*).

To support the RG picture and to describe the
"gapped" phases, we made use of  the Gaussian variational method (GVM) in
replica space, as introduced in \cite{parisi} and used in Ref.~\cite{GVM} to
treat interacting 1D disordered media. This method is able to capture
the localized compressible phases. To start with, we
rewrite the action $S_0$ in Fourier space as
\begin{equation}
S^0 = \frac{1}{2}\frac{1}{\beta L} \sum_{q,i\omega_n}\phi_\alpha^a(q,i\omega_n)(G_0^{-1})_{\alpha \beta}^{ab}(q,i\omega_n)\phi_\beta^b(-q,-i\omega_n),\nn
\end{equation}
where $\alpha,\beta$=$f,b$ while Latin indices run from 1 to $n$,
the number of replicas.
The free propagator is a $2n\times 2n$ matrix given by
$(G_0^{-1})_{\alpha \alpha}^{ab}(q,i\omega_n)$=$\delta_{ab}/(\pi
K_\alpha)(\omega_n^2/v_\alpha + v_\alpha q^2)$ and
$(G_0^{-1})_{fb}^{ab}(q,i\omega_n)=
(G_0^{-1})_{bf}^{ab}(q,i\omega_n)=\delta_{ab} U_{bf} q^2 /\pi^2 $.
The idea of the GVM is to replace the complicated action $S$ in
replica space by its best Gaussian approximation, $S_G$,
with $(G^{-1})_{\alpha \alpha}^{ab} = (G_0^{-1})_{\alpha \beta}^{ab}-\sigma_{\alpha \beta}^{ab}$, and $\sigma_{\alpha \beta}^{ab}$ the self-energy.
The optimal $G$ can then be obtained
   by minimizing the variational free energy,
$F_{\rm var}$=$F_G  +  \langle S-S_G\rangle_G/\beta$ with respect to $G_{\alpha
   \beta}^{ab}$.

\begin{figure}
\includegraphics[width=7.5cm]{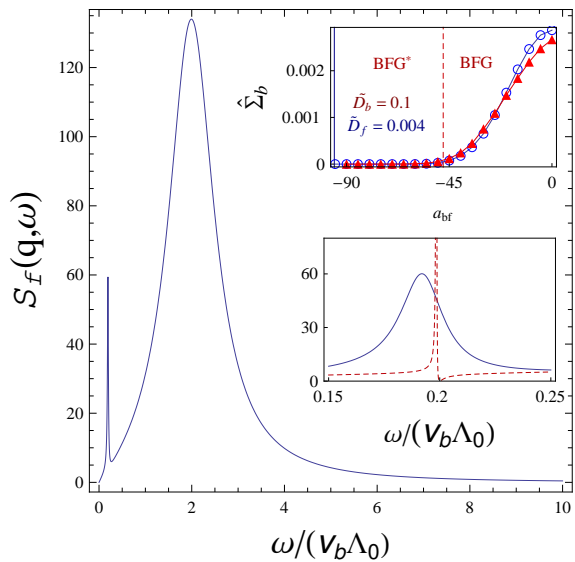}
\caption{Main plot: fermion structure factor
$S_f(q = 0.2\Lambda_0,\omega)$ in
  the BFG phase as computed by GVM. Bose-Fermi scattering length is $a_{bf}=-35a_0$. The UV cut-off is $\Lambda_0 = \rho_b$. We have taken
  $V_{\perp,b}=40 E_{R,b}$. Other parameters are the
  ones of Fig. \ref{Fig:phase_diag_exp}. Bottom inset is a zoom around
  the first peak at $\omega = v_b(q^2+\hat{\Sigma}_b)^{1/2}$, a signature of the coupled nature of the Bose-Fermi glass. The additional dashed curve is the structure factor in the AG+SFB phase ($V_{\perp,b}=13 E_{R,b}$).  Top inset shows
  $\hat{\Sigma}_b = \xi_b^{-2}$, as obtained from the RG
  flow (circles) and the 2-step RSB variational calculation (triangles). $V_{\perp,b}$ (i.e. $K_b$) is fixed and the boundaries
  obtained from the first RG analysis are represented. One can see that
  in region BFG$^*$, the mass is non zero but extremely small. Below
  $a_{bf}=-95a_0$ ($X_f=3$), $\hat{\Sigma}_b$ is identically zero.}
\label{Fig:main}
\end{figure}

Similar to Ref.~\cite{GVM},
we find that the phase with localized fermions and superfluid bosons
is well described by 
assuming one step replica symmetry breaking (1RSB) in the fermionic sector
\cite{future_paper}. We obtain a fermion mass $\hat{\Sigma}_f = \xi_f^{-2}$,
\beq
\hat{\Sigma}_f =
\left( \frac{K_f^2}{v_f^2}\frac{D_f}{\sqrt{1-g^2}}\frac{2}{\Lambda}\right)^{\frac{2}{3-X_f}}\Lambda^2,
\eeq
in agreement with the RG, and the solution ceases to exist when $X_f>
3$.  An RG calculation with respect to the latter variational action
($S_G$ with $\xi_f$ finite) provides a modified RG equation  which indeed smoothly interpolates between
the limiting cases in Eq.~\eqref{Db_scaling}, and can be used to
compute the  localization length, $\xi_b$. In the intermediate region
BFG$^*$ we find an extremely large, but finite  $\xi_b$
(see Fig.~\ref{Fig:main}). Similarly, we can compute  the superfluid stiffness
$\displaystyle \mathcal{D} = \lim_{\omega \rightarrow 0} \lim_{q
  \rightarrow 0} \omega^2 \overline{\langle \phi_b(q,\omega)
  \phi_b(-q,-\omega)\rangle}$ in the AG+SFB phase using our
variational solution. We find that  $\mathcal{D}>0$  only if
$K_b>3/2$,  indicating a phase transition
towards a Bose Glass phase. This BFG* phase  for $K_b<3/2$
appears in the variational problem as a
a level 2 replica-symmetry breaking solution (RSB), with the replica
symmetry also broken in the bosonic sector. 
However, $\xi_b$ being extremely large in the BFG$^*$
region, it may appear as a superfluid phase
in a finite system~\cite{future_paper}. {The nature of the transition between the BFG and the LL is quite subtle and will be detailed in  \cite{future_paper}}.

{\em Observables.} The dynamical response of the Bose-Fermi mixture
varies  from one phase to another. Such a response can
be probed through Bragg scattering measurements
(see {\it e.g.}~\cite{Inguscio_bragg}), giving access
to the dynamical structure factor 
$S_{b/f}({q},\omega) = \int dt dx \ e^{iqx-i\omega t }
\overline{\langle \rho_{b/f}(x,t)\rho_{b/f}(0,0)\rangle}$.
The variational approach enables us to compute $S_{b/f}({q},\omega)$, at
least for small values of $q$.
Fig.~\ref{Fig:main} shows { the inelastic part of} $S_f(q,\omega)$ computed
at two points of the phase
diagram. { The elastic part that comes from the gauge transformation used to absorb forward scattering would appear as a Dirac delta at $\omega$=0}.  The main plot shows the fermion structure factor in the BFG phase.
It exhibits a two peak structure with peak locations
 $\omega \approx v_b(q^2 + \hat{\Sigma}_b)^{1/2}$
and $\omega \approx v_f(q^2 + \hat{\Sigma}_f)^{1/2}$. The first peak is a bosonic feature, indicating that the Bose-Fermi glass is indeed a coupled localized phase.
The main effect of disorder is to introduce new
energy scales in the structure factor, $v_f \hat{\Sigma}_f^{1/2}$ and
$v_b \hat{\Sigma}_b^{1/2}$. Its other effect is to introduce a
linear frequency dependence for small $\omega$~\cite{GVM,future_paper}. 
Notice that in the localized phases there is no hard
gap in the excitation spectrum. The bottom inset is a zoom around the bosonic peak. Its counterpart in the AG+SFB phase is plotted there (for clarity it has been left out in the main plot). Note that in the AG+SFB
$\hat{\Sigma}_b$ is zero and the peak at $\omega \sim v_b
q$ is much sharper, as it would be in a simple Luttinger liquid. Finally,
in the LL phase, where both species are free, one would simply find two
very sharp peaks, corresponding to the two sound modes of the mixture,
at $\omega = v_+ q$ and $\omega = v_- q$, with $v_\pm^2 =
\frac{1}{2}(v_f^2 +
v_b^2)\pm\frac{1}{2}\left((v_f^2-v_b^2)^2+4g^2v_f^2v_b^2\right)^{1/2}$
\cite{cazalilla1,polaron1}. 

The transition from the  AG+SFB state to the BFG phase can
easily be detected through time of flight (TOF) measurements. In the
AG+SFB phase one has $\langle \psi_b(x)\psi^\dagger_b(x')\rangle \sim
1/|x-x'|^{1/2K_b^*}$, with $K_b^*$ the renormalized Luttinger parameter of
the bosons. As a result, a bosonic coherence peak is predicted
with a power law  dependence, $n_b(R)\sim 1/R^{1-1/2K_b^*}$, with
$n_b$ the density of bosons as measured in the TOF experiment,
and $R$ the expansion along the 1D tubes.

{\em Conclusion} To summarize, we have established the phase diagram
of an interacting Bose-Fermi mixture in the presence of uncorrelated
disorder. For relevant experimental parameters, in the case of a
$^{87}$Rb-$^{40}$K mixture, we have found three different phases,
including a Bose-Fermi glass where both species are coupled and
localized. These new phases can be detected by Bragg scattering and
time of flight measurements.

{\it Acknowledgment.} We acknowledge fruitful discussions with
N. Laflorencie.
This work has been partially supported
by the Institut Universitaire de France,
the OTKA Grant No. K73361, and the CNCSIS grant No. ID672/2009.

\vspace{-0.9cm}


\end{document}